\documentclass[runningheads]{llncs}
\usepackage[T1]{fontenc}

\usepackage{graphicx}
\usepackage{color}
\usepackage{amsmath,amssymb,amsfonts}
\usepackage{pifont}
\usepackage{multirow}
\usepackage{makecell}
\usepackage{bm}
\usepackage{booktabs}
\usepackage[misc]{ifsym}

\definecolor{myRed}{RGB}{195,10,10}
\definecolor{myGreen}{RGB}{55,149,73}

\newcommand{\cmark}{\textcolor{myGreen}{\ding{51}}}
\newcommand{\xmark}{\textcolor{myRed}{\ding{55}}}

\usepackage{hyperref}
\hypersetup{colorlinks = True, 
	    linkcolor = red, 
        citecolor = blue} 
\begin{document}
\title{FastSAM3D: An Efficient Segment Anything Model for 3D Volumetric Medical Images}
\titlerunning{FastSAM3D}
\author{
Yiqing Shen\inst{1} \and Jingxing Li\inst{1} \and Xinyuan Shao\inst{1} \and Blanca Inigo Romillo\inst{1} \and Ankush Jindal\inst{1} \and David Dreizin\inst{2}\textsuperscript{(\Letter)} \and Mathias Unberath\inst{1}\textsuperscript{(\Letter)}
}
\authorrunning{Y. Shen et al.}
\institute{
Johns Hopkins University, Baltimore, MD, 21218, USA \and
University of Maryland School of Medicine and R Adams Cowley Shock Trauma Center, Baltimore, MD, 21201, USA\\
\email{
\{yshen92,unberath\}@jhu.edu, daviddreizin@gmail.com}
}

\maketitle              
\begin{abstract}
Segment anything models (SAMs) are gaining attention for their zero-shot generalization capability in segmenting objects of unseen classes and in unseen domains when properly prompted. 
Interactivity is a key strength of SAMs, allowing users to iteratively provide prompts that specify objects of interest to refine outputs.
However, to realize the interactive use of SAMs for 3D medical imaging tasks, rapid inference times are necessary.
High memory requirements and long processing delays remain constraints that hinder the adoption of SAMs for this purpose.
Specifically, while 2D SAMs applied to 3D volumes contend with repetitive computation to process all slices independently, 3D SAMs suffer from an exponential increase in model parameters and FLOPS. 
To address these challenges, we present \texttt{FastSAM3D} which accelerates SAM inference to 8 milliseconds per $128\times128\times128$ 3D volumetric image on an NVIDIA A100 GPU. 
This speedup is accomplished through 
1) a novel layer-wise progressive distillation scheme that enables knowledge transfer from a complex 12-layer ViT-B to a lightweight 6-layer ViT-Tiny variant encoder without training from scratch; 
and 2) a novel 3D sparse flash attention to replace vanilla attention operators, substantially reducing memory needs and improving parallelization.
Experiments on three diverse datasets reveal that \texttt{FastSAM3D} achieves a remarkable speedup of $527.38\times$ compared to 2D SAMs and $8.75\times$ compared to 3D SAMs on the same volumes without significant performance decline.
Thus, \texttt{FastSAM3D} opens the door for low-cost truly interactive SAM-based 3D medical imaging segmentation with commonly used GPU hardware.
Code is available at \url{https://github.com/arcadelab/FastSAM3D}.

\end{abstract}

\section{Introduction}
In medical image analysis, object segmentation is a key aspect of diagnosis- and prognosis-related tasks including lesion localization, tissue characterization, and volume estimation, among others~\cite{dora2017state,mirikharaji2023survey,wu2022swin,jiang2022deep}.
Traditionally, deep learning models like U-Net \cite{UNet} and variants \cite{ke2023clusterseg,he2023transnuseg} have excelled in specific tasks and datasets with clear and confined scope, but often demonstrate limited generalization.
While some work considered interactive segmentation approaches as a means to overcome the limitations of narrowly scoped, task-specific models~\cite{amrehn2017ui,wang2018interactive}, the introduction of Segment Anything Model (\texttt{SAM})~\cite{sam} initiated a paradigm shift to prompt-based interactive segmentation that now provides competitive performance due to the inherent generalizability of foundation models.
%
\texttt{SAM} is comprised of a pre-trained Vision Transformer (ViT) encoder \cite{vit}, a prompt encoder, and a lightweight decoder that facilitates multi-mask prediction via IoU-based ranking. 
Trained on over 1 billion masks and 11 million images, \texttt{SAM} adapts to new tasks without training \cite{sam}.
Despite successes on natural images \cite{gao2024survey,liu2023towards}, direct application of \texttt{SAM} to medical segmentation reveals performance gaps compared to task-specific U-Nets \cite{zhang2024segment}.

\begin{table*}[!t]
\caption{
Comparison of SAM approaches regarding applicability for medical imaging, suitability for 3D volumetric data, and computational efficiency of the core components: image encoder, prompt encoder, and mask decoder.
The vanilla \texttt{SAM}~\cite{sam} lacks in all criteria. 
\texttt{MobileSAM}~\cite{mobilesam} improves encoder efficiency, while \texttt{TinySAM}~\cite{tinysam} accelerates all components, but neither addresses 3D medical imaging data.
\texttt{MedSAM}~\cite{medsam} and \texttt{SAM-Med2D}~\cite{sammed2d} are tailored for medical 2D data yet do not improve efficiency. 
\texttt{SAM-Med3D}~\cite{sammed3d} handles 3D medical data but inference times for these limit or altogether preclude real-time interactive use with standard GPU hardware.
Our proposed \texttt{FastSAM3D} meets all criteria, providing a comprehensive solution for efficient interactive medical image segmentation in volumetric 3D data.
}\label{table:feature_comparison}
\centering
\resizebox{\linewidth}{!}{
\begin{tabular}{l|cc|ccc} 
\toprule
Method & Medical & \makecell[c]{Volumetric\\ 3D Data} & \makecell[c]{Efficient \\Image Encoder} & \makecell[c]{Efficient \\Prompt Encoder} & \makecell[c]{Efficient \\ Mask Decoder} \\
\hline
\texttt{SAM}~\cite{sam} & \xmark & \xmark  & \xmark & \xmark & \xmark \\
\hline
\texttt{MobileSAM}~\cite{mobilesam} & \xmark & \xmark  & \cmark & \xmark & \xmark \\
\texttt{TinySAM}~\cite{tinysam} & \xmark & \xmark  & \cmark & \cmark & \cmark \\
\hline
\texttt{MedSAM}~\cite{medsam} & \cmark & \xmark  & \xmark & \xmark & \xmark \\
\texttt{SAM-Med2D}~\cite{sammed2d} & \cmark & \xmark  & \xmark & \xmark & \xmark \\
\texttt{SAM-Med3D}~\cite{sammed3d} & \cmark & \cmark  & \xmark & \xmark & \xmark \\
\hline
\texttt{FastSAM3D} (ours) & \cmark & \cmark  & \cmark & \cmark & \cmark \\
\bottomrule
\end{tabular}
}
\end{table*}

To address this, \texttt{MedSAM} \cite{medsam} and \texttt{SAM-Med2D} \cite{sammed2d} were tailored for 2D medical data via model fine-tuning.
When applied to 3D volumetric data, these approaches under-perform due to slice-wise processing~\cite{mazurowski2023segment,bui2023sam3d}.
They also suffer from an increased computational cost that is proportional to the number of slices in the volume, as well as the higher input resolution.
Addressing this gap, \texttt{SAM-Med3D}~\cite{sammed3d} introduced 3D counterparts of SAM’s components and end-to-end 3D training.

Existing medical SAMs also face limitations of long inference times and high computational costs stemming from the Transformer architecture~\cite{vit}.
Prior efforts accelerated 2D SAMs for natural images via approaches such as \texttt{FastSAM}, which employs a YOLOv8 as the image encoder~\cite{bolya2019yolact,fastsam}.
However, this CNN-based approach exhibits limitations with small object segmentation and deviates from SAM’s interactive prompting design~\cite{mobilesam,fastsam}.
Attempts, more aligned with the SAM value proposition, such as \texttt{MobileSAM}, retain the Transformer encoder while employing distillation to transition from a larger SAM encoder to a more lightweight ViT encoder~\cite{mobilesam}.
Following a similar approach, other works have explored several ViT variants as alternative encoders to balance efficiency and effectiveness~\cite{zhang2024efficientvit,wang2023repvit}. 
\texttt{TinySAM} further reduces computational load via post-distillation quantization~\cite{tinysam}.
However, these advancements remain confined to 2D natural images without delving into efficient volumetric medical segmentation.
Table~\ref{table:feature_comparison} offers a systematic comparison of existing SAMs, highlighting the capabilities of our proposed method in addressing the demands of 3D medical image segmentation with enhanced efficiency across all components.

The contributions of this work are two-fold, summarized as follows:
Firstly, we introduce \texttt{FastSAM3D}, a markedly more efficient 3D SAM for interactive volumetric medical image segmentation. 
Rather than costly training from scratch which also leads to the difficulty in convergence \cite{kd}, we propose a layer-wise progressive distillation approach to transfer representational knowledge from a complex 12-layer ViT-B architecture to an efficient customized 6-layer ViT-Tiny encoder.
This retains segmentation performance while significantly enhancing computational efficiency.
Secondly, we propose a novel 3D sparse flash attention that replaces the standard self-attention operator in all SAM components, dramatically reducing memory footprint, and enabling parallel processing.
Together, these innovations address the efficiency limitations that hinder the implementation of medical SAMs for real-time prompt-based interactive 3D segmentation.

\section{Methods}
\subsection{Architecture Overview of \texttt{FastSAM3D}}
\begin{figure}[t!]
    \centering
    \includegraphics[width=0.9\linewidth]{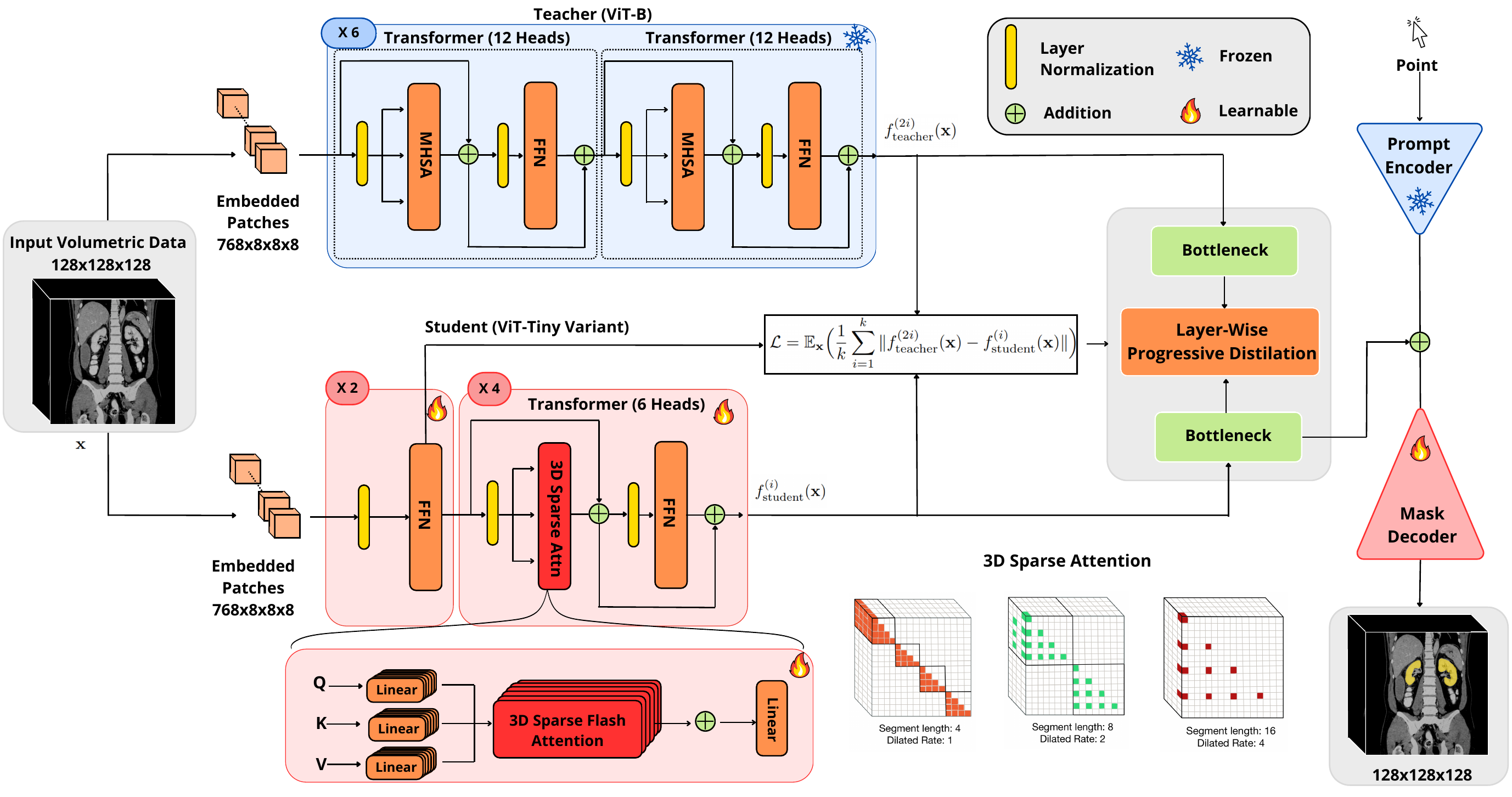}
    \caption{The overall framework of \texttt{FastSAM3D}, comprising a 6-layer ViT-Tiny variant image encoder distilled from a capable 12-layer ViT-B teacher encoder, a lightweight prompt encoder, and a mask decoder. 
    All the self attention operators are replaced by the proposed 3D sparse attention for better efficiency.
    }\label{fig:framework}
\end{figure}

We introduce \texttt{FastSAM3D}, a computationally efficient adaptation of \texttt{SAM-Med3D} \cite{sammed3d}, also designed specifically for efficient interactive 3D medical image segmentation.
Adhering to the standard SAM paradigm~\cite{sam}, \texttt{FastSAM3D} is comprised of three key modules (Fig.~\ref{fig:framework}): (i) a ViT-based image encoder \cite{vit} to obtain volumetric embeddings; (ii) a prompt encoder; and (iii) a mask decoder to project representations back to the segmentation mask.
To achieve faster inference, \texttt{FastSAM3D} distills knowledge from a high-powered 12-layer ViT-B encoder to a streamlined 6-layer ViT-Tiny variant, substantially reducing computational complexity during encoding.
Specifically, aside from having fewer layers, each Transformer block contains only 6 attention heads, in contrast to 12 heads per block in \texttt{SAM-Med3D}'s ViT architecture.
Moreover, we retain the feed-forward network (FFN) within the first two transformer blocks and omit attention operations~\cite{vit}, incurring minimal impact on performance while amplifying speed~\cite{yu2022metaformer}.
This design choice further contributes to shorter training by requiring fewer layers to align with the teacher during our progressive distillation process.

\subsection{Layer-wise Progressive Distillation for the Image Encoder}
As the image encoder accounts for a major portion of SAM’s computational load, our first focus is transferring knowledge from the heavy ViT-B architecture to a lightweight ViT-Tiny model for efficiency gains.
To avoid costly training from scratch, we follow the teacher-student distillation paradigm~\cite{kd} by designating the 12-layer ViT-B as the teacher model, $f_\text{teacher}$, and the 6-layer ViT-Tiny variant as the student model, $f_\text{student}$.
Unlike traditional logit-level distillation~\cite{kd} with which all our experiments failed to converge, we propose a novel layer-wise progressive distillation method.
This approach allows for a more granular and effective knowledge transfer between the student and teachers by matching the intermediate representation progressively across layers, thus making it easier for optimization.
Formally, let $f_\text{teacher}^{(i)}(\mathbf{x})$ and $f_\text{student}^{(j)}(\mathbf{x})$ denote layer $i=1,\cdots,12$ and $j=1,\cdots,6$ outputs for an input $\mathbf{x} \in \mathbb{R}^{128\times128\times128}$ from the 12-layer teacher and 6-layer student respectively.
The objective of our layer-wise progressive distillation becomes:
\begin{equation}
   \mathcal{L} = \mathbb{E}_{\mathbf{x}}\Big(  
    \frac{1}{k} \sum_{i=1}^k\|f_\text{teacher}^{(2i)}(\mathbf{x}) - f_\text{student}^{(i)} (\mathbf{x})\|
\Big),\label{eq:obj}
\end{equation}
where $\|\cdot\|$ denotes the L2-norm, $k$ varies from 1 to 6 based on current and total training iterations:
\begin{equation}
    k =  \lceil \frac{\# (\text{Current Iteration}) \times 6 }{\# (\text{Total Iterations})}  \rceil,
    \label{eq:k}
\end{equation}
where $\lceil\cdot\rceil$ is the upper rounding operator.
This enables progressive alignment of student and teacher intermediate representations.
After finishing layer-wise distillation, we perform logit-level distillation to fit predictions further.
In Eq.~\ref{eq:obj}, $\mathbb{E}(\cdot)$ represents the expectation over all possible images.

\subsection{3D Sparse Flash Attention}
As we observe that the attention operators take up the largest proportion of computation, we introduce a 3D sparse flash attention operator to further enhance efficiency. 
Specifically, our 3D sparse flash attention scheme supplants the traditional self-attention operation in both the encoder and decoder, integrating extended receptive fields inspired by dilated convolutions~\cite{yu2015multi} with the computational agility achieved by flash attention~\cite{flashattention,flashattention2}.

\subsubsection{3D Sparse Attention} 
The 3D sparse attention mechanism aims to expand the receptive field across volumetric data while effectively managing the computational load. 
Traditional attention mechanisms tend to escalate in computational demand proportional to the increase in data volume, particularly challenging for 3D volumetric data due to its large number of tokens~\cite{vit,movit}.
To address this, our approach segments the input token sequence into equally sized partitions of $w$ and applies a strategic sparsification across these segments~\cite{longnet}. 
This involves selectively sampling data points at the determined intervals with length $r$, thereby diminishing the overall number of tokens subjected to the attention process.
This allows for more efficient computation by focusing attention on fewer yet representative tokens.
Formally, the 3D sparse attention mechanism can be formulated as computing the attention over each segment as follows:
\begin{equation}
    \hat{S}_i = [S_{i}, S_{i+r}, \ldots, S_{i+(w-1)r}],
\end{equation}
where $\hat{S}_i$ represents the selectively sampled segment, ensuring that the model's attention is distributed across a sparse set of points, thereby reducing computational demands without sacrificing the depth of contextual analysis.

\subsubsection{Enhancing Efficiency through Parallel Processing with Flash Attention} 
We enhance efficiency by processing each segment in 3D sparse attention independently, enabling parallel operations that significantly boost computational throughput.
By incorporating flash attention~\cite{flashattention,flashattention2}, our model optimizes the functionality of parallel attention heads, substantially reducing the time and memory overhead associated with simultaneous processing activities.

\subsubsection{Overall Processing Procedure} 
The 3D sparse flash attention operator, integral to both the image encoder and mask decoder, operates through a sequence of orchestrated steps as follows. 
The process starts with the sparsification step, wherein the input sequence undergoes partitioning into sparse segments.
Subsequently, the attention operation ensues, wherein the previous segments are subjected to the flash attention for parallelization~\cite{flashattention,flashattention2}.
The focus of this stage is on harnessing the reduced sequential computation and memory optimization capabilities of flash attention.
The final phase is recomposition~\cite{longnet}, where the discrete outputs procured from the flash attention are reassembled to form the final encoded representation. 
This stage ensures that the final encoded representation has an identical dimension to its input.

\section{Experiments}

\begin{table*}[!t]
\caption{
Performance comparison of 2D and 3D \texttt{SAM} approaches in terms of Dice score.
We measure the performance at 1, 3, 5, and 10 point prompts (pt).
\texttt{SAM-Med3D} and our \texttt{FastSAM3D} are evaluated in a 3D context, whereas \texttt{SAM}, \texttt{MobileSAM}, \texttt{TinySAM}, \texttt{MedSAM} and \texttt{SAM-Med2D} are applied independently to all 2D slices of the entire 3D volume.
Notably, \texttt{FastSAM3D} demonstrates competitive performance with \texttt{SAM-Med3D} and shows enhanced Dice scores relative to all its 2D counterparts, highlighting the effectiveness of our approach.
The best performance is shown in \textcolor{red}{\textbf{red}} and boldface, while the second best is in \textcolor{blue}{blue}.
}\label{table:performance}
\centering
\resizebox{\linewidth}{!}{
\renewcommand{\arraystretch}{1.6}
\begin{tabular}{c|c|c|c|c|c|c|c|c|c|c|c|c|c} 
\toprule
\multirow{2}{*}{Dim} & \multirow{2}{*}{Method} & \multicolumn{4}{c|}{AMOS~\cite{amos}} & \multicolumn{4}{c|}{TotalSegmentator~\cite{totalsegmentator}} & \multicolumn{4}{c}{BraTS~\cite{brats}}\\
\cline{3-14}
& & 1pt & 3pt & 5pt & 10pt & 1pt & 3pt & 5pt & 10pt & 1pt & 3pt & 5pt & 10pt \\
\hline
\multirow{5}{*}{2D} & \texttt{SAM}~\cite{sam} & $0.049$ & $0.093$ & $0.114$ & $0.145$ & $0.202$ & $0.279$ & $0.311$ & $0.348$ & $0.108$ & $0.192$ & $0.217$ & $0.237$ \\
\cline{2-14}
& \texttt{MobileSAM}~\cite{mobilesam} 
& $0.041$ & $0.056$ & $0.063$ & $0.070$ & $0.149$ & $0.170$ & $0.182$ & $0.212$ & $0.079$ & $0.132$ & $0.156$ & $0.186$ \\
\cline{2-14}
& \texttt{TinySAM}~\cite{tinysam}
& $0.049$ & $0.077$ & $0.089$ & $0.101$ & $0.171$ & $0.225$ & $0.243$ & $0.262$ & $0.103$ & $0.165$ & $0.187$ & $0.211$\\
\cline{2-14}
& \texttt{MedSAM}~\cite{medsam}
&$0.004$ & $0.051$ & $0.060$ & $0.074$ & $0.006$ & $0.069$ & $0.090$ & $0.111$ & $0.008$ & $0.059$ & $0.064$ & $0.071$ \\
\cline{2-14}
& \texttt{SAM-Med2D}~\cite{sammed2d}
&$0.097$ & $0.127$ & $0.129$ & $0.132$ & $0.008$ &$0.081$ & $0.100$& $0.128$ & $0.013$ &$0.076$ & $0.082$ & $0.084$ \\
\hline
\multirow{2}{*}{3D} & \texttt{SAM-Med3D}~\cite{sammed3d} & \textcolor{red}{\bm{$0.289$}} & \textcolor{red}{\bm{$0.386$}} & \textcolor{red}{\bm{$0.418$}} & \textcolor{red}{\bm{$0.448$}} & \textcolor{red}{\bm{$0.252$}} & \textcolor{red}{\bm{$0.400$}} & \textcolor{red}{\bm{$0.463$}} & \textcolor{red}{\bm{$0.522$}} & \textcolor{blue}{$0.328$} & \textcolor{blue}{$0.395$} & \textcolor{blue}{$0.418$} & \textcolor{red}{\bm{$0.446$}}\\
\cline{2-14}
& \texttt{FastSAM3D} & \textcolor{blue}{$0.273$} & \textcolor{blue}{$0.368$} & \textcolor{blue}{$0.402$} & \textcolor{blue}{$0.437$} & \textcolor{blue}{$0.250$} & \textcolor{blue}{$0.378$} & \textcolor{blue}{$0.445$} & \textcolor{blue}{$0.519$} & \textcolor{red}{\bm{$0.333$}} & \textcolor{red}{\bm{$0.401$}} & \textcolor{red}{\bm{$0.421$}} & \textcolor{blue}{$0.445$}\\
\bottomrule
\end{tabular}
}
\end{table*}

\subsubsection{Implementation Details}
Our method as well as all baseline methods are implemented in Python 3.9 and PyTorch 2.1.0.
The computational environment for our experiments is standardized across all methods, utilizing an NVIDIA A100 GPU with 40Gb of memory.
For the layer-wise progressive distillation, we set the total training iteration number in Eq.~\eqref{eq:k} to $36$.
Training is facilitated by the Adam optimizer, with a learning rate of $5\times10^{-3}$ and a batch size of 16.
For evaluation metrics, we use the Dice score to measure segmentation performance.
We also report inference time, floating point operations (FLOPs), and memory cost to quantify the computational complexity.

\subsubsection{Datasets}
Our evaluation incorporates three diverse datasets that span two modalities, namely computed tomography (CT) and magnetic resonance imaging (MRI), where we follow the dataset splits of previous work~\cite{sammed3d}.
(1) The \textit{AMOS} dataset~\cite{amos} is a substantial and varied clinical collection designed for abdominal organ segmentation with 500 CT and 100 MRI scans.
(2) The \textit{TotalSegmentator} dataset~\cite{totalsegmentator} consists of 1228 CT studies each with 117 anatomical structures acquired from different pathologies, scanners, series, and institutions.
(3) The \textit{BraTS 2021} dataset~\cite{brats} assembles a total number of 1251 multi-institutional MRI scans.

\begin{table*}[!t]
\caption{
Comparison of the computational efficiency with respect to the encoder and decoder.
We report the time (ms), FLOPs (G), and memory (Gb), alongside acceleration factors relative to 2D \texttt{SAM}~\cite{sam} and 3D \texttt{SAM-Med3D}~\cite{sammed3d}. 
For 2D SAMs, we compute the time to process all the slices within volumetric data.
The best results are highlighted in \textbf{bold} if statistically different from the second best result ($p < 0.01$).
}\label{table:computation}
\centering
\resizebox{\linewidth}{!}{
\begin{tabular}{c|c|c|ccc|ccc|c|c} 
\toprule
\multirow{3}{*}{Dim} & \multirow{3}{*}{Method} & \multirow{3}{*}{Resolution} & \multicolumn{3}{c|}{Encoder} & \multicolumn{3}{c|}{Decoder} & \multicolumn{2}{c}{Acceleration} \\
\cline{4-11}
&&& \makecell[c]{Time\\(ms)$\downarrow$} & \makecell[c]{FLOPs\\(G)$\downarrow$} & \makecell[c]{Memory\\(Gb)$\downarrow$} & \makecell[c]{Time\\(ms)$\downarrow$} & \makecell[c]{FLOPs\\(G)$\downarrow$} & \makecell[c]{Memory\\(Gb)$\downarrow$} & 
To 2D $\uparrow$ & To 3D $\uparrow$ \\
\hline
\multirow{5}{*}{2D} & \texttt{SAM}~\cite{sam} & $1024\times1024$ & $3980$ & $369.0$ & $7.87$ & $239$ & $3.0$ & $5.57$ & $1.00\times$ & / \\
\cline{2-11}
& \texttt{MobileSAM}~\cite{mobilesam} &
$1024\times1024$ & $584$ & $36.7$ & $5.48$ & $233$ &  $3.0$ & $5.27$ & $5.16\times$ & /  \\
\cline{2-11}
& \texttt{TinySAM}~\cite{tinysam} & $1024\times1024$ & $609$ & $36.7$ & $5.48$ & $246$ & $3.0$ & $5.27$ & $4.93\times$ & / \\
\cline{2-11}
& \texttt{MedSAM}~\cite{medsam} & $1024\times1024$ & $3983$ & $369.0$ & $7.87$ & $241$ & $2.9$ & $5.57$ & $1.00\times$ & /\\
\cline{2-11}
& \texttt{SAM-Med2D}~\cite{sammed2d} & $256\times256$ & $1063$ & $32.0$ & $6.32$ & $216$ & \bm{$0.21$} & $5.55$ & $3.30\times$ & / \\
\hline
\multirow{2}{*}{3D} & \texttt{SAM-Med3D}~\cite{sammed3d} & $128\times 128\times128$ & $70$ & $89.5$ & $6.58$ & $20$ & $2.8$ & $5.53$ & $60.27\times$ & $1.00\times$ \\
\cline{2-11}
& \texttt{FastSAM3D}& $128\times 128\times128$ & \bm{$3$} & \bm{$21.9$} & \bm{$0.78$} & \bm{$5$} & $2.8$ & \bm{$0.71$} & \bm{$527.38\times$} & \bm{$8.75\times$} 
\\
\bottomrule
\end{tabular}
}
\end{table*}

\begin{table*}[!b]
\caption{
Ablation study for the contribution of 3D sparse attention (`Sparse Attn.') and flash attention (`Flash Attn.') to the performance and efficiency of \texttt{FastSAM3D}.
Best scores are highlighted in \textbf{bold}, if statistically different from the second best result ($p < 0.01$).
3D sparse attention and flash attention contribute to substantial improvements in time and memory requirements without statistically significant performance decline.
}\label{table:ablation}
\centering
\resizebox{\linewidth}{!}{
\begin{tabular}{c|c|cccc|cccc|cccc|ccc} 
\toprule
\multirow{2}{*}{\makecell[c]{Sparse\\ Attn.}} & \multirow{2}{*}{\makecell[c]{Flash\\Attn.}} & \multicolumn{4}{c|}{AMOS~\cite{amos}} & \multicolumn{4}{c|}{TotalSegmentator~\cite{totalsegmentator}} & \multicolumn{4}{c|}{BraTS~\cite{brats}} & \multicolumn{3}{c}{Encoder}\\
\cline{3-17}
& & 1pt & 3pt & 5pt & 10pt & 1pt & 3pt & 5pt & 10pt & 1pt & 3pt & 5pt & 10pt & Time & FLOPs & Memory \\
\hline
\xmark & \xmark & $0.282$ & {$0.375$} & {$0.403$} & $0.436$ & $0.243$ & $0.371$ & $0.442$ & $0.516$ & {$0.335$} & {$0.404$} & {$0.422$} & $0.444$ & $10$ & $23.1$ & $1.16$\\
\xmark & \cmark & $0.276$ & $0.366$ & $0.398$ & $0.432$ & $0.247$ & $0.374$ & $0.438$ & $0.516$ & $0.331$ & $0.402$ & $0.421$ & {$0.445$} & $6$ & \bm{$21.9$} & $1.15$\\
\cmark & \xmark & $0.277$ & $0.370$ & $0.402$ & $0.433$ & {$0.255$} & {$0.381$} & {$0.450$} & {$0.520$} & $0.328$ & $0.403$ & {$0.422$} & {$0.445$} & $9$ & $23.1$ & $0.79$\\
\cmark & \cmark & $0.273$ & $0.368$ & $0.402$ & {$0.437$} & $0.250$ & $0.378$ & $0.445$ & $0.519$ & $0.333$ & $0.401$ & $0.421$ & {$0.445$} & \bm{$3$} & \bm{$21.9$} & \bm{$0.78$}\\
\bottomrule
\end{tabular}
}
\end{table*}

\subsubsection{Performance Comparison}
\begin{figure}[t!]
    \centering
    \includegraphics[width=\linewidth]{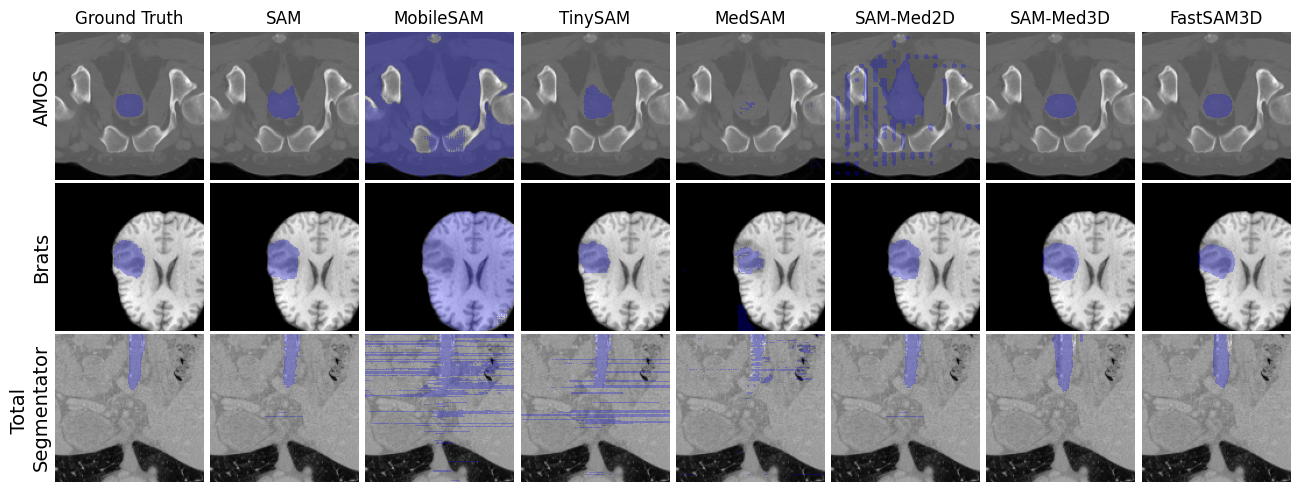}
    \caption{Representative segmentation results from all methods across three datasets. 
    \texttt{FastSAM3D} demonstrates accurate contour delineation comparable to \texttt{SAM-Med3D}.
    }\label{fig:result}
\end{figure}

Table~\ref{table:performance} compares segmentation performance for \texttt{FastSAM3D} with various 2D and 3D SAM approaches.
Fig.~\ref{fig:result} provides an illustrative visualization for samples segmented by different methods.
\texttt{FastSAM3D} not only demonstrates competitive segmentation performance in comparison to its teacher model, \texttt{SAM-Med3D}, but also surpasses all 2D efficient SAM models, especially when the number of point prompts is increased.
For example, \texttt{FastSAM3D} achieves a Dice score of $0.437$ on the \textit{AMOS} dataset with 10 point prompts, which is a significant improvement over the $0.306$ score from the best-performing 2D model ($p<0.01$). 
This trend is consistent across the \textit{TotalSegmentator} and \textit{BraTS} datasets, underscoring the robustness of \texttt{FastSAM3D} across different datasets, modalities, and organs. 
Additionally, 2D SAM methods require intensive per-slice prompting as opposed to \texttt{FastSAM3D} which only involves volume-level interactions.

\subsubsection{Computational Efficiency Comparison}
Regarding the computational efficiency, Table~\ref{table:computation} reveals that \texttt{FastSAM3D} reduces the inference time for the encoder to 3 milliseconds and decoder to 5 milliseconds for 3D volumetric images, a substantial improvement from the $3980$ milliseconds required by the vanilla \texttt{SAM} employed in a slice-by-slice manner.
Moreover, \texttt{FastMed3D} requires fewer FLOPs and less memory than all counterparts, achieving a $527.38\times$ acceleration compared to the vanilla \texttt{SAM} and $8.75\times$ acceleration compared to \texttt{SAM-Med3D}.

\subsubsection{Ablation Study}
Table~\ref{table:ablation} illustrates the effectiveness of both sparse and flash attention in optimizing computational efficiency.
Specifically, when neither 3D sparse nor flash attention mechanisms were applied, the model achieved a Dice score of $0.436$ on the AMOS dataset with 10 prompts. 
The introduction of 3D sparse attention marginally reduces the Dice score to $0.433$ from $0.436$ on \textit{AMOS} but substantially reduces memory consumption from $1.06$ Gb to $0.79$ Gb. 
Flash attention alone improves inference time from $10$ ms to $6$ ms, underscoring its impact on computational efficiency.
Moreover, the concurrent implementation of both sparse and flash attention yields the most substantial improvements. 
For instance, the Dice score on the AMOS dataset with 10 prompts increases to $0.437$, and the encoder time is reduced to 3 ms,
Memory requirements are also minimized to $0.78$ Gb, suggesting an optimized model footprint.

\section{Conclusion}
We present \texttt{FastSAM3D}, an innovative adaptation of \texttt{SAM} for efficient segmentation of volumetric medical imaging data. 
This model addresses the critical challenges of high inference time and the substantial computational cost associated with previous 3D \texttt{SAM} methods. 
Through a novel layer-wise progressive distillation and 3D sparse flash attention integration, we significantly reduce computational demands while maintaining high segmentation performance.
Our experiments across different modalities and organs demonstrate that \texttt{FastSAM3D} not only accelerates inference by factors of $527.38\times$ compared to 2D \texttt{SAM}s and $8.75\times$ to 3D \texttt{SAM}s but also retains the flexibility of SAM's interactivity, making it a promising and powerful tool for clinical deployment. 
\texttt{FastSAM3D} opens up the possibility of real-time human-machine interaction by facilitating rapid prompting volumetric segmentation, thereby potentially maximizing user agency and trust while minimizing effort, workload, and wait time-related frustration.
With the speed and efficiency of \texttt{FastSAM3D}, another possible direction includes the development of mixed reality (MR) applications for surgical planning and guidance.

\subsubsection{Acknowledgements.} This work was supported in part by grants from the National Institutes of Health (NIH R01 GM148987-01).

\bibliographystyle{splncs04}
\bibliography{main.bib}

\end{document}